\documentclass[aps,prl,groupedaddress,fleqn,twocolumn,10pt]{revtex4}
\pdfoutput=1
\usepackage{graphicx}
\usepackage{amsmath}
\usepackage{gensymb}
\usepackage{float}
\usepackage{physics}
\usepackage{xcolor}

\begin{document}

\raggedbottom

\title{The Purcell question: why do all viscosities stop at the same place?}
\author{K. Trachenko}
\affiliation{School of Physics and Astronomy, Queen Mary University of London, Mile End Road, London, E1 4NS, UK}
\author{V. V. Brazhkin}
\affiliation{Institute for High Pressure Physics, RAS, 108840, Troitsk, Moscow, Russia}

\begin{abstract}
In 1977, Purcell asked why liquid viscosities all stop at the same place? 
Liquids are hard to understand, yet today we can answer the Purcell question in terms of fundamental physical constants fixing viscosity minima. With the Planck constant setting the minimal viscosity, water and life appear to be well attuned to the degree of quantumness of the physical world.

\end{abstract}

\maketitle

In 1977, Purcell observed that there is almost no liquid with viscosity much lower than that of water and wrote in the first paragraph of Ref. \cite{purcell} (original italics preserved):\\

``The viscosities have a big range {\it but they stop at the same place}. I don't understand that''.\\

In the first footnote of that paper, Purcell says that Weisskopf has explained this to him. We did not find published Weisskopf's explanation, however the same year Weisskopf published the paper ``About liquids'' \cite{weiskopf}. That paper starts with a story often recited by conference speakers: imagine a group of isolated theoretical physicists trying to deduce the states of matter using quantum mechanics only. They are able to predict the existence of gases and solids, but not liquids.

Weisskopf implies that liquids are hard. This is hammered home in textbooks. For example, Landau and Lifshitz repeatedly assert \cite{landau} that contrary to gases and solids, liquid thermodynamic properties and their temperature dependence can not be calculated in general form due to the combination of strong inter-molecular interactions and the absence of small oscillations simplifying solid theory. This is the famous ``no small parameter'' problem.

A small parameter can still be found in liquids, but it is hidden in the liquid flow. In the theory of J. Frenkel, liquid molecules vibrate at short time (shorter than liquid relaxation time) like in solids and diffuse at long times only. Phonon vibrations give the small parameter as in solids but in liquids this process is short-lived. In other words, liquids have a smaller parameter but you need to be quick to see it. The consequences of this important fact have been seen in recent experiments and molecular modelling and are used to construct a theory of liquid thermodynamics. Tests of these theories are becoming increasingly detailed and rigorous \cite{proctor}.

Meanwhile, do we understand viscosity any better since 1977 to answer the Purcell question of why all viscosities ``stop at the same place''?

We know that viscosity $\eta$ of a dilute gas-like fluid is $\eta=\frac{1}{3}\rho v L$, where $\rho$, $v$ and $L$ are density, average velocity of molecules and mean free path, respectively. This equation predicts that $\eta$ of gases does not depend on density or pressure because $L\propto\frac{1}{\rho}$. It also predicts that gas viscosity {\it increases} with temperature because $v$ increases. This increase might be counter-intuitive because we often experience fluids getting thinner when heated. Think about bitumen flowing in hot weather or jam getting thicker when cooled in the fridge after cooking. Indeed, viscosity of dense liquids {\it decreases} with temperature as $\eta=\eta_0\exp\left(\frac{U}{T}\right)$, where $U$ is activation energy.


The increase of $\eta$ at high temperature and its decrease at low imply that $\eta$ has a {\it minimum}, and we expect this minimum to correspond to the crossover between the gas-like and liquid-like viscosity. It is convenient to look at this crossover above the critical point where it is smooth and no liquid-gas phase transition intervenes. Fig. 1 shows experimental kinematic viscosity $\nu=\frac{\eta}{\rho}$, where $\rho$ is density, of several supercritical fluids. $\nu$, as well as $\eta$, indeed have the minima, and we can understand them as the crossover between gas-like and liquid-like states.

\begin{figure}
\begin{center}
{\scalebox{0.38}{\includegraphics{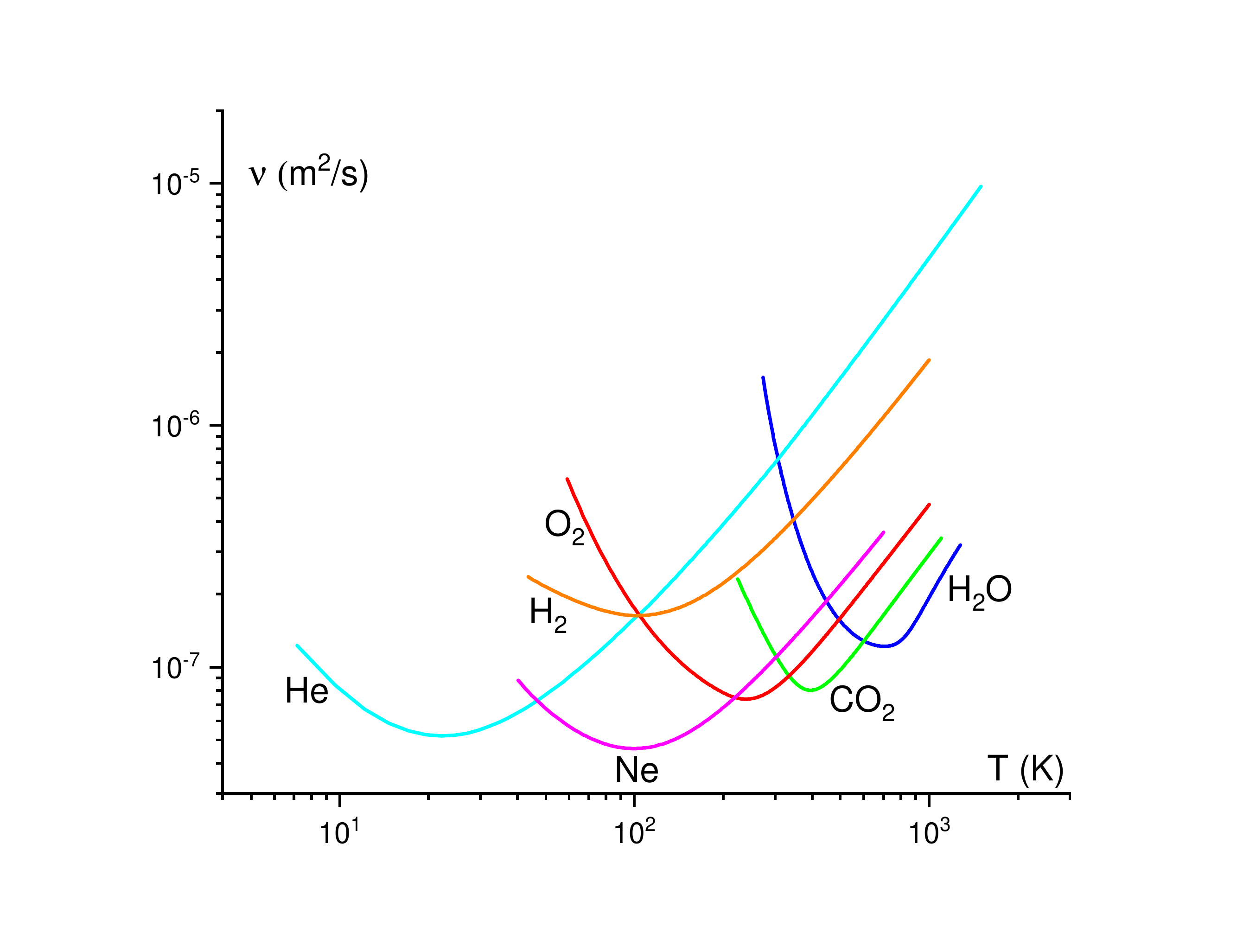}}}
\end{center}
\caption{Experimental kinematic viscosity of noble, molecular and network fluids showing minima. $\nu$ for He, H$_2$, O$_2$, Ne, CO$_2$ and H$_2$O are shown at 20 MPa, 50 MPa, 30 MPa, 50 MPa, 30 MPa and 100 MPa, respectively. The experimental data are from the National Institute of Standards and Technology database, see https://webbook.nist.gov/chemistry/fluid.}
\label{3}
\end{figure}

The viscosity minima provide the first clue to the Purcell question: viscosities stop decreasing due to a minimum. But can the minimum itself get arbitrarily close to zero? (disclaimer: we do not discuss superfluidity here). Why the minima of $\eta$ are hard to move up or down and are somehow close to the viscosity of water at room conditions? We may be able to answer this if we can calculate viscosity at the minimum.

Calculating viscosity is hard. The reason is the same as in the Landau and Lifshitz textbook: interactions in liquids are strong and system-specific. It is hard to calculate viscosity parameters even in simpler fluids using theory only and no input from modelling. For molecular and network liquids like H$_2$O, this is unworkable. 

It turns out that the minimum of viscosity at the crossover represents a very special point where viscosity can in fact be evaluated, even though approximately. More specifically, it turns out that the minima of kinematic viscosity $\nu$, $\nu_m$, can be related to only two basic properties of a condensed matter system: interatomic separation and Debye vibration frequency - both of them representing a UV cutoff in the system. These two parameters can be, in turn, related to the Bohr radius $a_{\rm B}=\frac{\hbar^2}{m_ee^2}$ ($m_e$ is the electron mass) setting a characteristic interatomic distance on the order of Angstroms and the Rydberg energy giving a characteristic bonding energy. Then, $\nu_m$ becomes \cite{sciadv}

\begin{equation}
\nu_m=\frac{1}{4\pi}\frac{\hbar}{\sqrt{m_em}}
\label{nu0}
\end{equation}

\noindent where $m$ is the molecule mass.

Two fundamental constants, $\hbar$ and $m_e$, appear in Eq. \eqref{nu0}. The minimal viscosity turns out to be quantum! This may seem surprising and at odds with our thinking about high-temperature liquids as classical. Eq. \eqref{nu0} reminds us that the nature of interactions in condensed matter is quantum-mechanical in origin, with $\hbar$ affecting the Bohr radius and Rydberg energy.

The fundamental constants help keep $\nu_m$ from moving up or down too much. $\nu_m$ are not universal due to $\nu_m\propto\frac{1}{\sqrt{m}}$ mass dependence, although this does change $\nu_m$ too much for most liquids. We are getting close to answering the Purcell question.

For different fluids such as those in Fig. 1, Eq. (\ref{nu0}) predicts $\nu_m$ in the range (0.3-1.5)$\cdot 10^{-7}\frac{{\rm m}^2}{\rm s}$. This fairs well with experiments \cite{sciadv} and is somewhat lower, but not far, from $\nu$ in water at room conditions. Water at room temperature happens to be runny enough and close to the minimum. This is what Purcell noted: viscosities of most liquids do not go much lower than in water.

So the answer to the Purcell question has two parts: (a) viscosities ``stop'' because they have minima and (b) the minima are fairly fixed by fundamental constants.

Incidentally, the same happens to an unrelated liquid property: thermal diffusivity. It also ``stops'' at the same place and, interestingly, is given by the same Eq. (\ref{nu0}). This implies that the transfer of both momentum and energy at the minimum involves the same mechanism.

We can now introduce two viscosities which are truly universal. The first is the fundamental kinematic viscosity obtained from Eq. (\ref{nu0}) by setting $m=m_p$, where $m_p$ is the proton mass, corresponding to atomic H:

\begin{equation}
\nu_f=\frac{1}{4\pi}\frac{\hbar}{\sqrt{m_em_p}}\approx 10^{-7}\frac{\rm m^2}{\rm s}
\label{nu1}
\end{equation}

The second property is the ``elementary'' viscosity $\iota$ (``iota'') defined as $\iota=\nu_m m$. Using \eqref{nu0}, we see that $\iota$ has the absolute lower bound, $\iota_m$, for $m=m_p$ as

\begin{equation}
\iota_m=\frac{\hbar}{4\pi}\left({\frac{m_p}{m_e}}\right)^{\frac{1}{2}}\approx 3\hbar
\label{iota}
\end{equation}

\begin{figure}
\begin{center}
{\scalebox{0.25}{\includegraphics{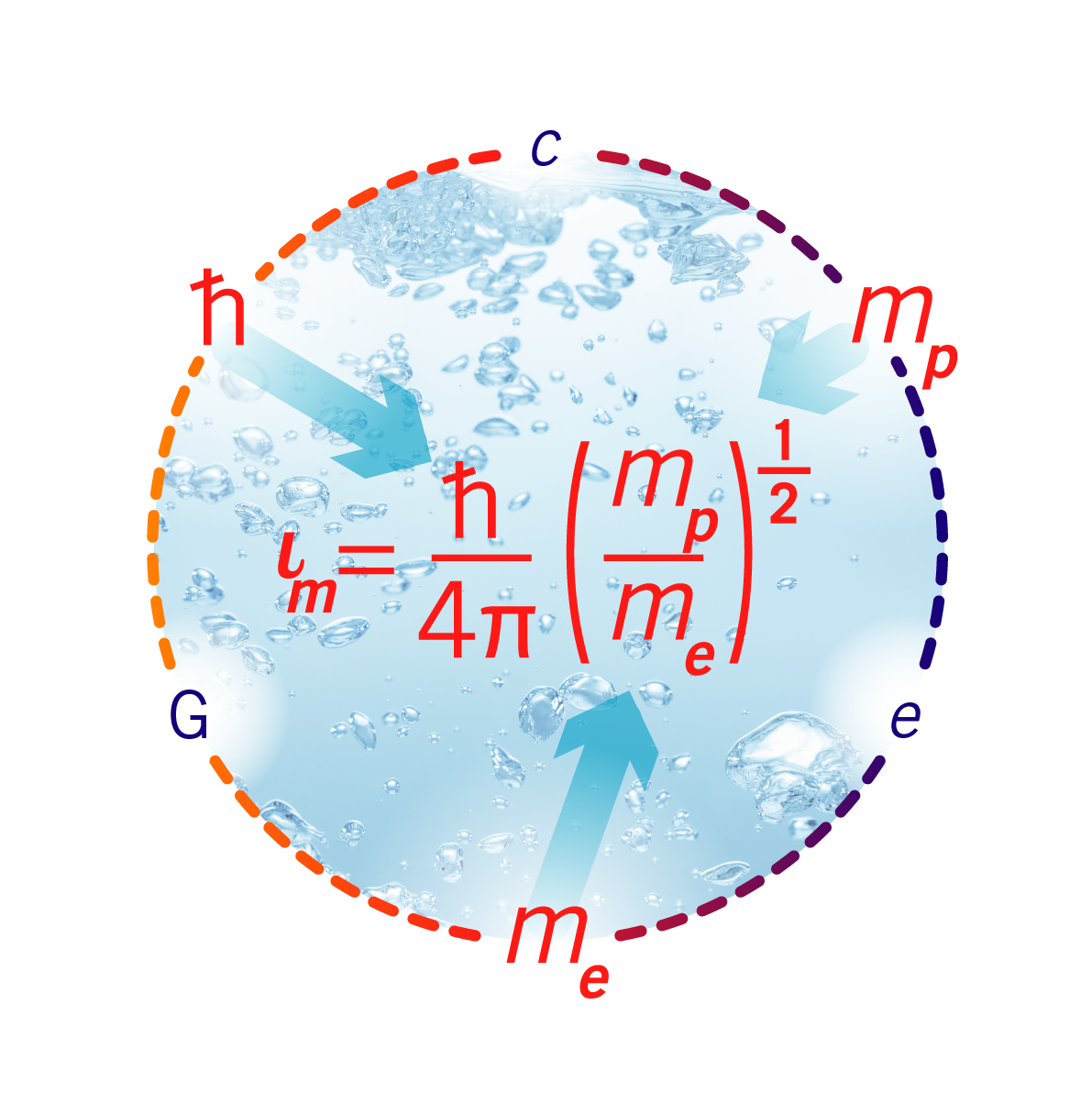}}}
\end{center}
\caption{Three fundamental constants setting the minimal elementary viscosity.}
\label{3}
\end{figure}

Eq. \eqref{iota}, illustrated in Fig. 2, interestingly contains the proton-to-electron mass ratio, one of three dimensionless fundamental constants of general importance. Together with two other dimensionless constants, it decides whether our Universe is bio-friendly by affecting the formation of stars, synthesis of heavier elements including carbon and ordered molecular structures \cite{barrow}.

These are all profound happenings. We can add another observation: the fundamental constants are friendly to life at a higher level too: biological processes in cells rely heavily on water. Were fundamental constants (say, $\hbar$) to take different values, water viscosity would change too, both kinematic viscosity $\nu$ relevant for water flow and dynamic viscosity $\eta$ setting internal friction, diffusion and so on. $\nu$ and $\eta$ at the minimum depend on $\hbar$ differently: whereas $\nu_m\propto\hbar$ in \eqref{nu0}, the minimum of dynamic viscosity, $\eta_m$, is very sensitive to $\hbar$: $\eta_m=\nu_m\rho\propto\frac{\nu_m}{a_{\rm B}^3}\propto\frac{\nu_m}{\hbar^6}\propto\frac{1}{\hbar^5}$. If the viscosity minimum increases due to a different $\hbar$ and water becomes too viscous as a result, processes in cells would not be the same. Life might not exist in its current form or not exist at all. One might hope that cells could still survive in such a Universe by finding a hotter place where overly viscous and bio-unfriendly water is thinned. This would not help though: $\hbar$ sets the minimum below which viscosity can not fall regardless of temperature. This way, water and life are well attuned to the degree of quantumness of the physical world.


\end{document}